# Estimate of the Mass Composition of Ultrahigh Energy Cosmic Rays

A.A. Mikhailov, N.N. Efremov, N.S. Gerasimova, I.T. Makarov,
G.V. Nikolayeva, G.G. Struchkov.
*Yu.G. Shafer Institute of Cosmophysical Research and Aeronomy, 31 Lenin Ave., 677980 Yakutsk, Russia*
Presenter: A. Mikhailov (mikhailov@ikfia.ysn.ru), rus-mikhailov-AA-abs1-HE14-oral

It is proposed a new approach for estimating the composition of cosmic rays. It is found that the zenith angle distributions and muon components of EAS' for energies $E>10^{19}$ eV and $E>4\times10^{19}$ eV differ from each other. It is shown that the cosmic rays above $E>4\times10^{19}$ eV is heavier than the cosmic rays at energy $E\sim 10^{19}$ eV. According to our estimation the SUGAR array detected 8 showers above $10^{20}$ eV. It is concluded that no sign of Greisen-Zatsepin-Kuz'min (GZK) cut off in the spectrum of cosmic rays and all cosmic rays are galactic.

## 1. Introduction

The composition of cosmic rays is the important characteristic to solve a problem of their origin. To clarify this question, the muon shower component as the most sensitive to the change of primary cosmic ray composition can play the essential role. The analysis of the muon component of extensive air showers (EAS') by using AGASA array data (Japan) shows that in cosmic rays at $E>10^{19}$eV the light nuclei are dominated [1]. The results obtained at the Hires array (USA) by data of the shift rate of shower development maximum depending on the energy show that cosmic rays at $E\sim2.5\times10^{19}$eV consist of light nuclei, most likely [2]. The estimation of cosmic ray composition at the Yakutsk EAS array by the Cherenkov radiation points to the fact that cosmic rays at $E\sim3\times10^{19}$eV consist mainly of the protons also [3]. Unfortunately, in these papers to interpret experimental data the model calculations are used which consider NN – and $\pi$N – interactions of very high-energy particles whose cross-sections are extrapolated from the accelerator region. In this extrapolation the inaccuracies can be. The experiments are also difficult and errors are not excluded.

Here we propose a new approach for estimating the composition of cosmic ray on the basis of clearly determined experimental data.

## 2. Discussion

Fig.1 presents the distribution of EAS' with $E>10^{19}$eV in zenith angle $\theta$: a - Yakutsk, b – Haverah Park [4]. The number of showers is 458 and 144, respectively. The dashed line is the expected number of events on the observation level according to [5]. Pearson $\chi^2$ – criterion shows that between observed and expected numbers of showers there is the good agreement at a significance level of 0.05. As seen in Fig.1, in the shower distribution with $E>10^{19}$eV the inclined showers are predominated.

In Fig.2 the EAS distribution at $E>4\times10^{19}$eV is shown: a – Yakutsk, b – AGASA [6]. The number of EAS' is equal to 29 and 47. The dashed line is the expected number of events on the observation level. For Yakutsk array the observed number of EAS' does not contradict the expected number of EAS' according $\chi^2$ – criterion at a significance level of 0.05. The same is observed on the data of array AGASA (fig. 2b). If to unit these two distributions of showers (Yakutsk and AGASA) that the observed number of EAS' contradict the expected number of events at a significance level of 0.05. At that in an interval of angles 20° - 30° the observed number of EAS' exceeds the expected ones on 2.3σ, where σ - standard deviation from expected number of events.

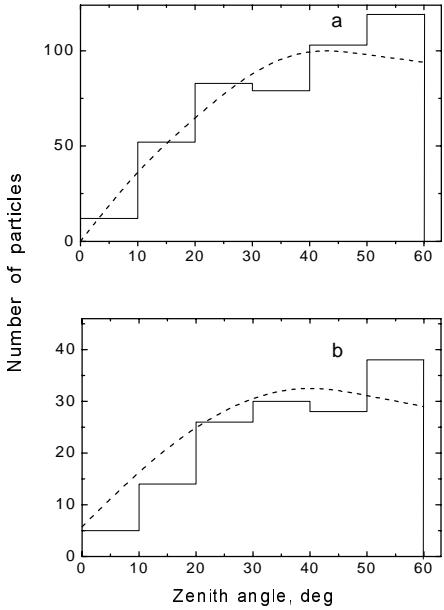

Fig.1. Distribution of showers with E>10$^{19}$ eV in zenith angle θ: a-Yakutsk, b-Haverah Park. The dashed line is the expected number of showers on the observation level.

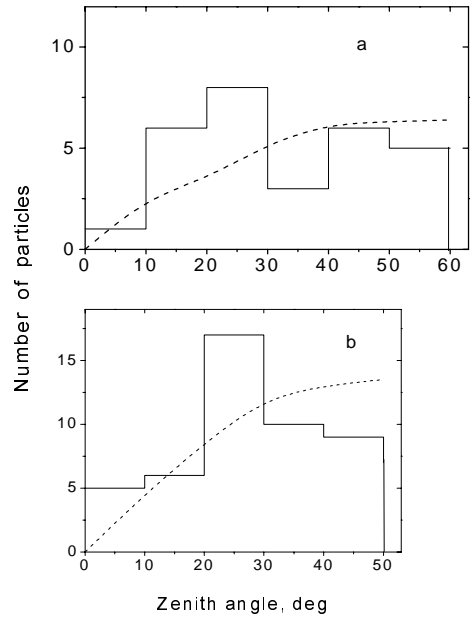

Fig.2. Distribution of EAS' with E>4×10$^{19}$ eV: a-Yakutsk, b-AGASA. The dashed line is the expected number of showers on the observation level.

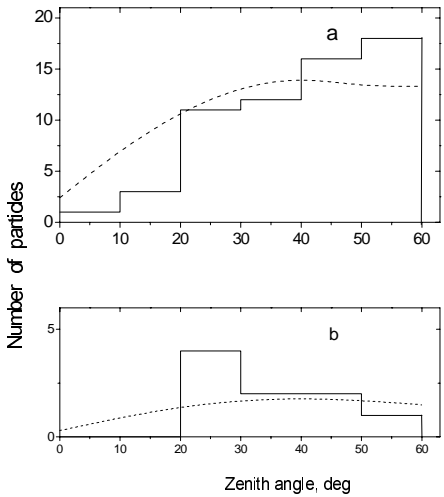

Fig.3. Distribution of showers detected at the Sugar array with E>10$^{19}$eV (a) and 4×10$^{19}$eV (b) according to the "Sydney" model and (a) with E>4×10$^{19}$eV according to the "Hillas-E" model. The dashed line is the expected number of showers on the observation level.

Thus, the shower distribution in the zenith angle at E>10$^{19}$eV and E>4×10$^{19}$eV differs from each other.

We consider the EAS distribution in zenith angle by the SUGAR data. In [7] there are two variants to estimate of the shower energy: by the "Sydney" model and the "Hillas - E" model.

Fig.3 shows the shower distributions according the "Sydney" model: the showers with E>10$^{19}$eV (a) and with E>4×10$^{19}$eV (b) but among them there are no showers with E>10$^{20}$ eV. The EAS' distribution in the zenith angle at E>4×10$^{19}$eV contradict the expected number of EAS'. Obviously the estimation of energy EAS' by the "Sydney" model is not correct.

According to the "Hillas – E" model, the showers in Fig.3a have the energies higher than 4×10$^{19}$eV. The shower distribution in zenith angle (Fig.3a) is agreed according χ$^2$ – criterion the expected number of EAS'.

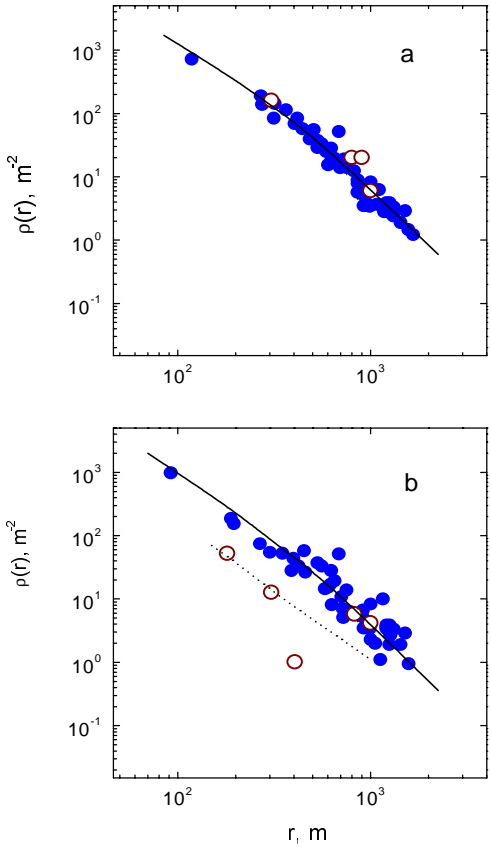

Fig.4. Particle density ρ(r) versus the distance r to a shower core: a-$E_1=1.2\times10^{20}$eV, b-$E_2=2\times10^{19}$eV, ●-electrons and photons, ○-muons, the solid and dashed lines are the expected densities of the electron-photon component and muons.

On this basic, one can conclude that the estimation of the shower energy by the "Hillas – E" model is more correct, or according to this model at E>$10^{20}$eV 8 EAS' are registered [7]. The SUGAR data ("Hillas – E" model) support AGASA results that no sign GZK cut off in the spectrum of cosmic rays.

Possible in Yakutsk and Haverah Park arrays the energy of EAS' at E>$4\times10^{19}$eV decrease and in results it is observed at E>$10^{19}$eV and θ>50° more EAS' than expected one (Fig.1). We suppose that energy EAS' at Yakutsk and AGASA arrays at E>$4\times10^{19}$eV must be increase.

In order to clarify why the zenith angle distribution of EAS' contradict to expected ones at E>$10^{19}$ eV and E>$4\times10^{19}$ eV (Fig.1, 2), we consider these showers by Yakutsk data.

Fig.4, demonstrates as an example of all data, the electron-proton and muon components of two inclined showers with angles and energies: a-$\theta_1=58.7°$, $E_1=1.2\times10^{20}$eV; b-$\theta_2=54.5°$ and $E_2=2\times10^{19}$eV. These showers are registered on May 7, 1989 and December 2, 1996 at the Yakutsk EAS array. The axes of the two showers are inside the array perimeter. As seen in Fig.4a, the particle densities in the scintillation detectors (registration threshold of electrons and photons is 3 MeV) and in the 4 muon detectors (threshold is 1 GeV) become equal, i.e. the shower with $E_1=1.2\times10^{20}$eV consists of muons only. The shower with $E_2=2\times10^{19}$eV at the same zenith angle θ has the electron-photon and muon components (Fig.4b). Why the electron-photon component of EAS' disappear at E ~ $10^{20}$eV ? The fact that a portion of muons in the inclined showers at E>$10^{19}$ eV increases and electron-photon disappear with the energy is established over all data in [8].

Thus, two facts have been established: 1) the zenith angle distribution of EAS' at E>$10^{19}$eV and E>$4\times10^{19}$eV differs from each other, 2) the muon component of EAS' at E>$10^{19}$eV beginning to increase. This facts can be interpreted as the change of the mass composition of cosmic rays between at E=(1-4)$\times10^{19}$eV to the side of more heavy nuclei.

The qualitative picture of the shower development is: a heavy nucleus interacts with air atoms in relatively high layer of the atmosphere in comparison with more light nuclei and disintegrates on the nucleons. In result are create the showers of small energy, and the electron-photon component of EAS' in relatively smaller energies are apparently absorbed stronger. Therefore a deficit of electron-photon component in inclined EAS' takes place (Fig.4a). On this basis it may be concluded that the mass composition of cosmic rays with E>$4\times10^{19}$eV is more heavy than cosmic rays at E~$10^{19}$eV - iron nuclei Fe. Earlier we showed that cosmic rays at E~$10^{19}$eV are most likely the iron nuclei [9]. Note that we showed that the cosmic rays at E>$4\times10^{19}$eV correlate with pulsars of the Local Arm of Galaxy [10,11] and see also [12].

## 3. Conclusions

Cosmic rays with $E>4\times10^{19}$ eV are more heavy than the iron nuclei Fe and galactic.

## 4. Acknowledgements

The work has been supported by RFBR (project N 04-02-16287). The Yakutsk EAS array was supported by the Ministry of Training of the Russian Federation, project no.01-30.